\documentclass[aip,amsmath,amssymb,reprint]{revtex4-1}

\draft % marks overfull lines with a black rule on the right

\usepackage{graphicx}% Include figure files
\usepackage{dcolumn}% Align table columns on decimal point
\usepackage{bm}% bold math
\usepackage[utf8]{inputenc}
\usepackage[T1]{fontenc}
\usepackage{mathptmx}

\begin{document}

%\preprint{}

\title{Free-standing 2D metals from binary metal alloys} %Title of paper



\author{Janne Nevalaita}
\author{Pekka Koskinen}
\email[]{pekka.j.koskinen@jyu.fi}
%\thanks{}
\affiliation{Department of Physics, Nanoscience Center, University of Jyv\"{a}skyl\"{a}, 40014 Jyv\"{a}skyl\"{a}, Finland}


\date{\today}

\begin{abstract}
Recent experiment demonstrated the formation of free-standing Au monolayers by exposing Au--Ag alloy to electron beam irradiation. Inspired by this discovery, we used semi-empirical effective medium theory simulations to investigate monolayer formation in $30$ different binary metal alloys composed of late d-series metals Ni, Cu, Pd, Ag, Pt, and Au. In qualitative agreement with the experiment, we find that the beam energy required to dealloy Ag atoms from Au--Ag alloy is smaller than the energy required to break the dealloyed Au monolayer. Our simulations suggest that similar method could also be used to form Au monolayers from Au--Cu alloy and Pt monolayers from Pt--Cu, Pt--Ni, and Pt--Pd alloys.
\end{abstract}


\maketitle %\maketitle must follow title, authors, abstract and \pacs


Common two-dimensional (2D) materials have a layered bulk structure, where covalently bonded layers are held together by van der Waals forces~\cite{miro14,lin16,novoselov19}, enabling monolayer exfoliation~\cite{coleman11,huang15}. However, recent experiments have discovered 2D materials with non-layered bulk geometries, such as transmission electron microscopy observations of 2D iron patches inside graphene nanopores~\cite{zhao14}. Computational studies motivated by this discovery have since predicted stable 2D metal monolayers composed of elements beyond Fe~\cite{nevalaita18, nevalaita18b}, including Au, Ag, and Cu~\cite{yang15b,yang15,yang16}. Besides their importance for fundamental research, these free-standing metal atom monolayers have numerous potential applications, including catalysis and sensing~\cite{chen18}.

However, isotropic bonding in metals renders conventional fabrication methods inapplicable to manufacturing free-standing metal atom monolayers. An alternative to layer exfoliation is solid-melt exfoliation, used to fabricate Ga monolayers on multiple substrates~\cite{kochat18}, although not as free-standing monolayers. Recent experiments have taken a different approach. Transition metal dichalcogenide materials have a van der Waals structure and monolayers can be exfoliated in the conventional manner or fabricated by the novel conversion method from non-van der Waals solids~\cite{du2020}. The non-metal atoms can then be removed using electron irradiation. This technique has been used to fabricate free-standing Mo monolayers inside MoSe$_2$ template~\cite{zhao18}. Recently Wang and co-workers used this technique to selectively remove Ag atoms from Au--Ag alloy, leaving free-standing monolayers of Au atoms inside the Au--Ag alloy template~\cite{wang19}. Given this experimental proof-of-principle and the vast number of possible metal alloys, a theoretical study on other alloy candidates for the fabrication of monolayers is warranted.

Therefore, in this work, we explore the stability of monolayers in promising binary metal alloys using classical atomistic simulations. In our recent work~\cite{nevalaita19}, we concluded that the intrinsic stability of elemental metal monolayers is the greatest near the end of the d-series, which implies that the most promising binary metal alloys compose of Ni, Cu, Pd, Ag, Pt, and Au. Here we consider both the energetic and kinetic stability of resulting 30 binary combinations. We confirm that the experimentally observed dealloying of Ag atoms is kinetically feasible and that Au monolayers are stable when exposed to electron irradiation. Further, we predict that Au--Cu alloy could support stable monolayers of Au and Pt--Cu, Pt--Ni, and Pt--Pd alloys could support stable monolayers of Pt.
\begin{center}
\begin{figure}
\includegraphics[width=8.0 cm]{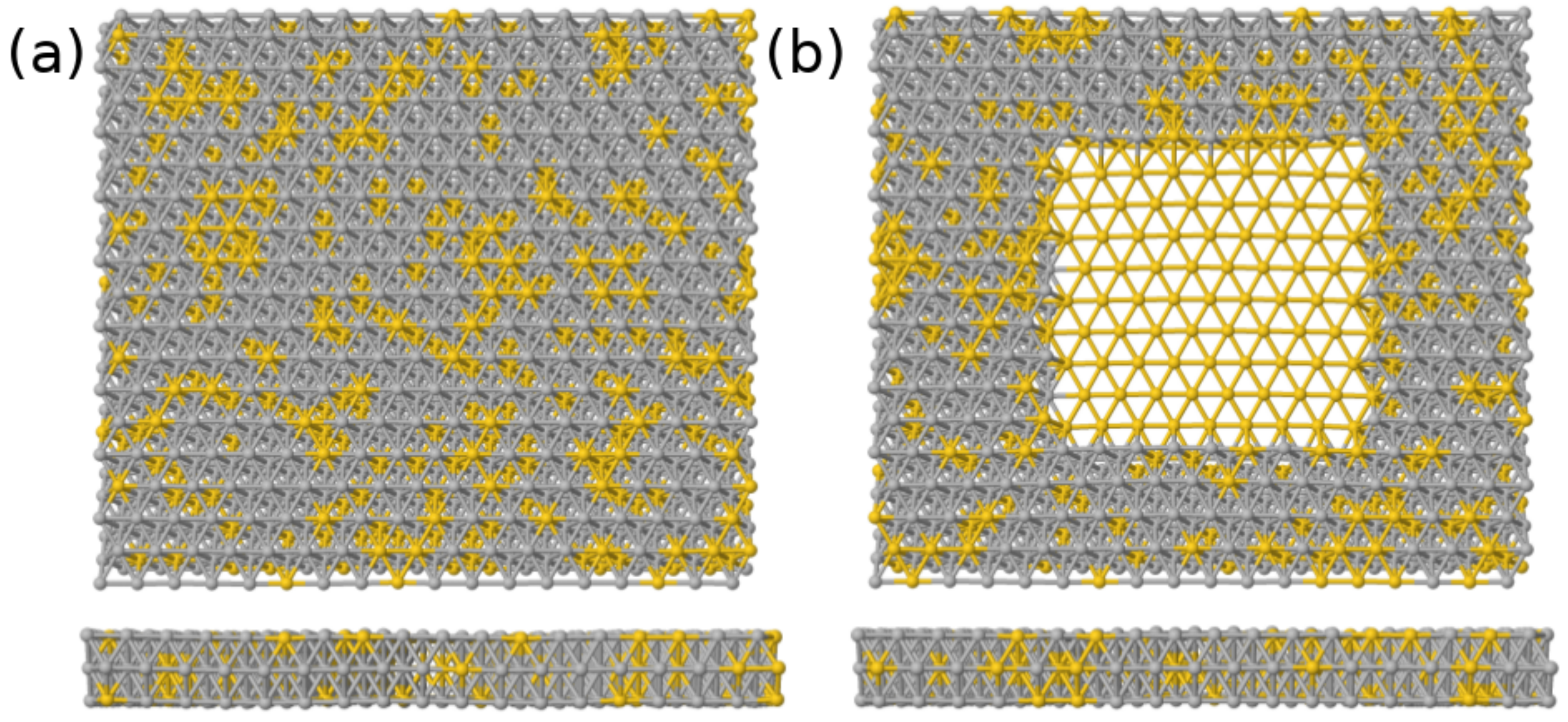}
\caption{Sketches of the simulated structures (top and side views). (a) Structures used in the simulation of alloy formation and dealloying energies. (b) Structures used in the simulations of monolayer stabilities. The alloy forms a template that consists of \emph{primary} atoms (here yellow spheres, Au) forming the monolayer and \emph{secondary} atoms (here grey spheres, Ag), which have supposedly been removed by a focused electron beam.}
\label{fig:sketch}
\end{figure}
\end{center}

To establish a common ground for kinetic studies, let us begin by considering static energetics of all 30 binary alloys. The alloys are simulated using the Atomic Simulation Environment (ASE)~\cite{ase-paper} and semi-empirical effective medium theory (EMT) with parameters given in Ref.~\onlinecite{jacobsen1996}. Although providing limited accuracy, EMT suffices well for our purposes as it usually preserves trends~\cite{jacobsen1996} and has been used to calculate low free energy structures of Au clusters with results comparable to density functional theory~\cite{garden18} as well as a starting point for determining the chemical ordering of Au-Ag clusters~\cite{larsen18}. The alloy formation energies are calculated from three-layer thick slabs with 972 atoms, where metal atoms are mixed randomly (Fig.~\ref{fig:sketch}a). The cells and atomic position are relaxed below force tolerance of $0.05$ eV/\AA. The Au--Ag dealloying experiment showed that the irradiation modifies the alloy structures to an extent that renders the exact (initial) atomic positions irrelevant.\cite{wang19}

\begin{center}
\begin{figure}
\includegraphics[width=8.0 cm]{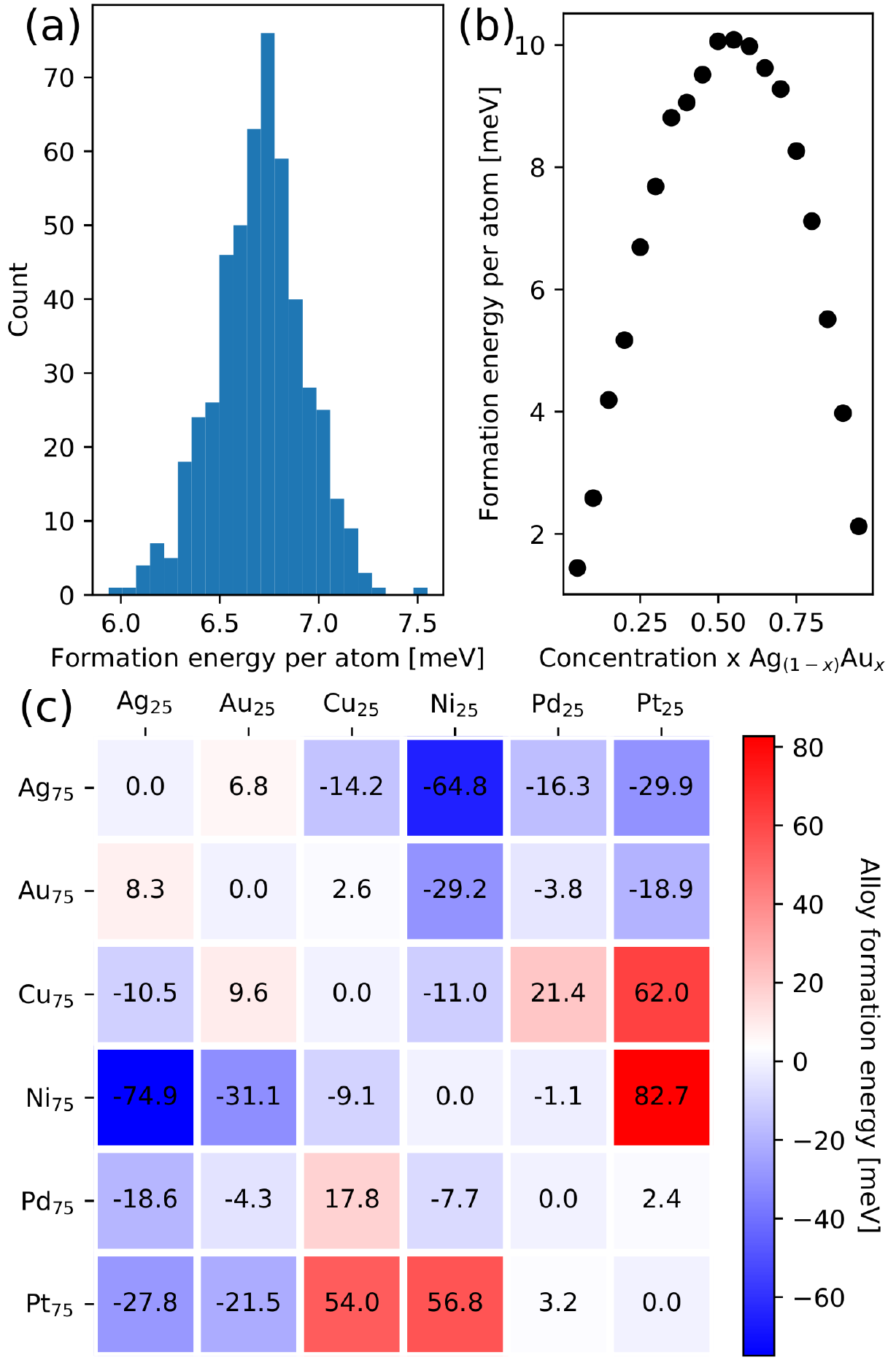}
\caption{Energetics of static alloys. (a) Alloy formation energies given by Eq.~\eqref{eq:Hf} for 500 Au$_{25}$Ag$_{75}$ alloys with random Au locations. The distribution has an average of 6.70 meV, standard deviation of 0.23 meV, and variance of 0.05 meV. (b) Formation energy as a function of Au concentration for Au--Ag alloy. (c) Alloy formation energies for 30 binary combinations. Positive energy (red color) indicates exothermic alloy formation.}
\label{fig:histo}
\end{figure}
\end{center}

To estimate the energetic stability of the alloys, we define the alloy formation energy as
\begin{equation}
\Delta H_f = E_{AB}-xE_A-(1-x)E_B,
\label{eq:Hf}
\end{equation}
where $x\in[0,1]$ is the $A$ concentration and $E_{A}$, and $E_{B}$ are the cohesive energies of metals $A$ and $B$. Here the cohesive energy $E_{AB}$ for the alloy is
\begin{equation}
E_{AB} = x\epsilon_A+(1-x)\epsilon_B-\epsilon_{AB},
\end{equation}
where $\epsilon_A$ and $\epsilon_B$ are the energies of free A and B metal atoms, and $\epsilon_{AB}$ is the energy per atom for the alloy. To justify the random alloying, we estimate its effect by calculating the formation energies of 500 Au$_{25}$Ag$_{75}$ alloys, composed of 25 \% of Au and 75 \% of Ag with random mixing (Fig. \ref{fig:histo}a). As expected, the formation energies are approximately normally distributed. The energy distribution averages to 6.7 meV with standard deviation of 0.23 meV. Such a narrow distribution indicates that the exact atomic positions are irrelevant for the formation energy. However, the alloy concentration itself has a substantial effect. While the concentration at macroscale can be controlled, at microscale it has local variations. Therefore, we consider the effect of concentration to the formation energy of Au--Ag alloy by calculating the energies for systems with varying ratio of Au and Ag atoms (Fig. \ref{fig:histo}b). Note that relatively small changes in the ratios between Au and Ag atoms changes the formation energy considerably. However, the previous experiment observed monolayer formation starting with Au concentrations between 15--35 \%~\cite{wang19}, indicating that the concentration range is sufficiently wide for feasible monolayer formation. We therefore focus here on alloys with 25 \% of primary atoms.

The formation energies for all binary combinations of Ni, Cu, Pd, Ag, Pt, and Au reveal both exothermic and endothermic alloying (Fig.~\ref{fig:histo}c). Formation energies are sign-wise symmetrical across the diagonal, indicating that if the alloy formation is exothermic at concentration $x = 0.25$ it is exothermic also at concentration $x = 0.75$. While many alloys have exothermic formation energies, only Au provides exothermic alloying for Ag. Similarly for Ni, only Pt alloying is exothermic.

\begin{center}
\begin{figure}[t!]
\includegraphics[width=8.6 cm]{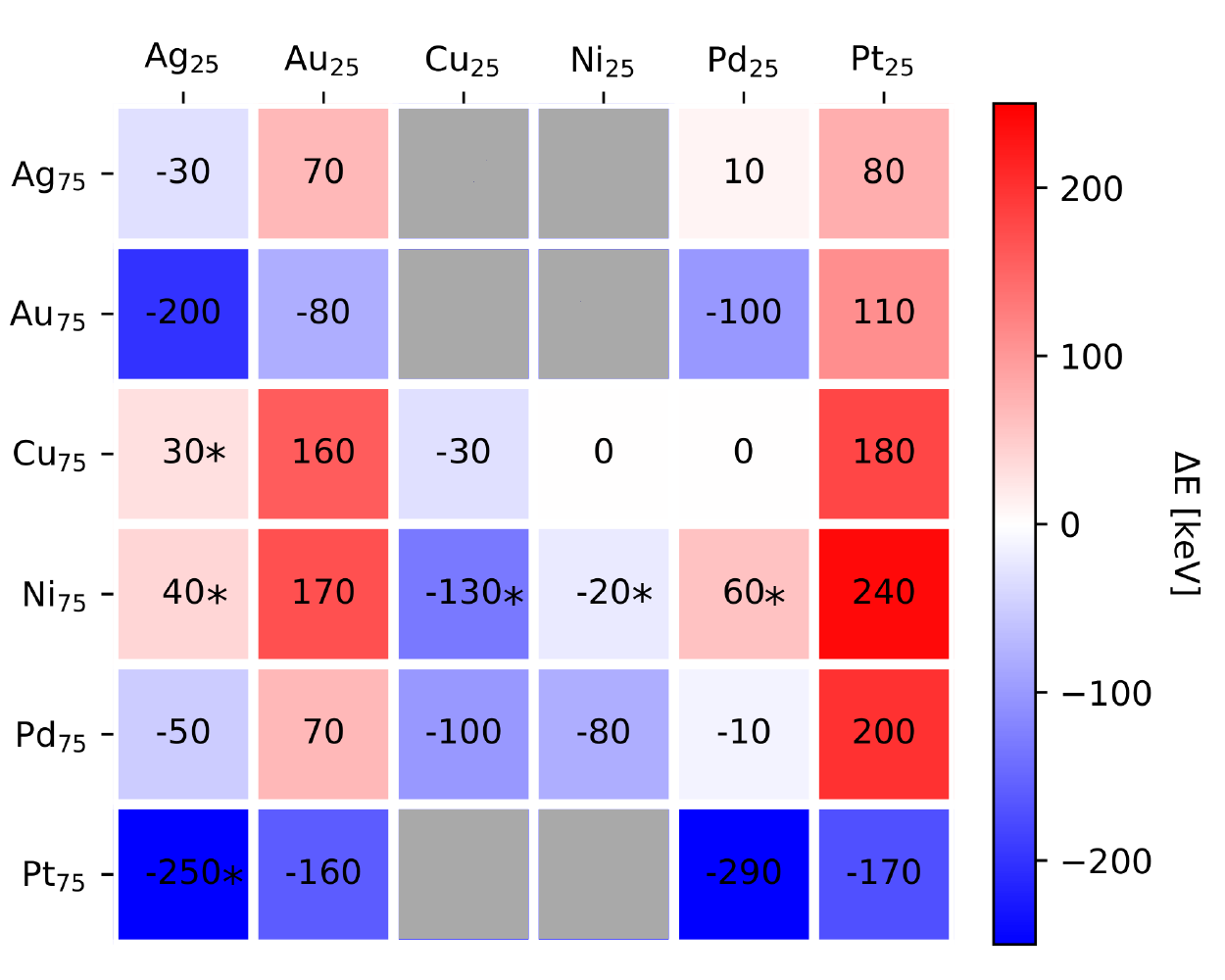}
\caption{Kinetic feasibility of monolayer formation. The array shows the window $\Delta E=E_\text{stable}-E_\text{dealloy}$, the energy difference between the maximum beam energy for a stable monolayer (principal metals, in columns) and minimum beam energy required to dealloy the secondary metal (in rows). Grey boxes indicate alloys for which monolayer breaks during adiabatic relaxation and values marked with asterisks indicate alloys for which the monolayer at room temperature is unstable.}
\label{fig:DE}
\end{figure}
\end{center}

To proceed from these somewhat familiar results toward kinetic effects, let us consider model systems with monolayers of primary metal within templates of binary alloy, irradiated by electron beam of kinetic energy $E$ (Fig. \ref{fig:sketch}b). To estimate the stability of a monolayer under electron irradiation, we run molecular dynamics simulations where an atom near the center of the monolayer was given a momentum perpendicular to the monolayer plane, mimicking the momentum transfer from a colliding electron, speeding at relativistic velocities. This momentum transfer could be coupled to beam energy $E$ via\cite{banhart1999}
\begin{equation}
T = \frac{2ME(E+2mc^2)}{(M+m)^2c^2+2ME},
\label{eq:Emax}
\end{equation} 
where $T$ is the kinetic energy given to the atom, $M$ the atom mass, $m$ the electron mass, and $c$ the speed of light. The momentum was increased until the monolayer got broken or the atom with initial momentum was displaced from its original position. Eq.~\eqref{eq:Emax} could then be used to obtain the maximum beam energy $E_\text{stable}$ tolerated by the monolayer. Some monolayers broke adiabatically during atomic relaxation and therefore lack values for maximum beam energies (grey squares in Fig.~\ref{fig:DE}). Further, the Cu monolayer in Cu$_{25}$Ni$_{75}$ alloy broke already at $T=0.1$ eV.

After examining the kinetic stabilities of monolayers, we calculate the electron beam energy $E_\text{dealloy}$ required to dealloy secondary atoms from the binary metal alloys by providing a metal atom at the bottom of a trilayer a momentum perpendicular to the trilayer. The momentum is increased until the atom escapes from its original position and the corresponding energy is associated with the electron beam energy via Eq.~(\ref{eq:Emax}). Since the elements in trilayers are randomly mixed, extractions of 20 randomly chosen atoms are calculated and their average $E_\text{dealloy}$ is used. The monolayer is kinetically feasible if there exists a window of electron beam energies where \emph{i}) the secondary metal is removed,  \emph{ii}) the primary metal is not removed, and \emph{iii}) the monolayer of the primary metal remains stable. Because in practice the energy to remove the primary metal is always larger than the stability limit of the monolayer, the sufficient condition for kinetic feasibility is reduced to the requirement of a positive dealloying energy window, given by $\Delta E=E_\text{stable}-E_\text{dealloy}>0$. Calculating these energy windows for all alloys reveals several binary metal alloy candidates that could be kinetically feasible for monolayer formation (positive numbers in Fig.~\ref{fig:DE}). 

Since the experimental Au monolayers are stable at room temperature, we tested the temperature stability of the model systems. We did this by molecular dynamics simulations, heating the alloys to 300 K using Langevin thermostat. As the result, we found that although some monolayers tolerate substantial momentum given to single atoms, they break upon heating to 300 K (alloys with asterisks in Fig.~\ref{fig:DE}). Among all binary alloys, only Au and Pt monolayers are stable at room temperature with all considered secondary atoms.\cite{Koskinen2015,Antikainen2017}

\begin{center}
\begin{figure}
\includegraphics[width=8.6 cm]{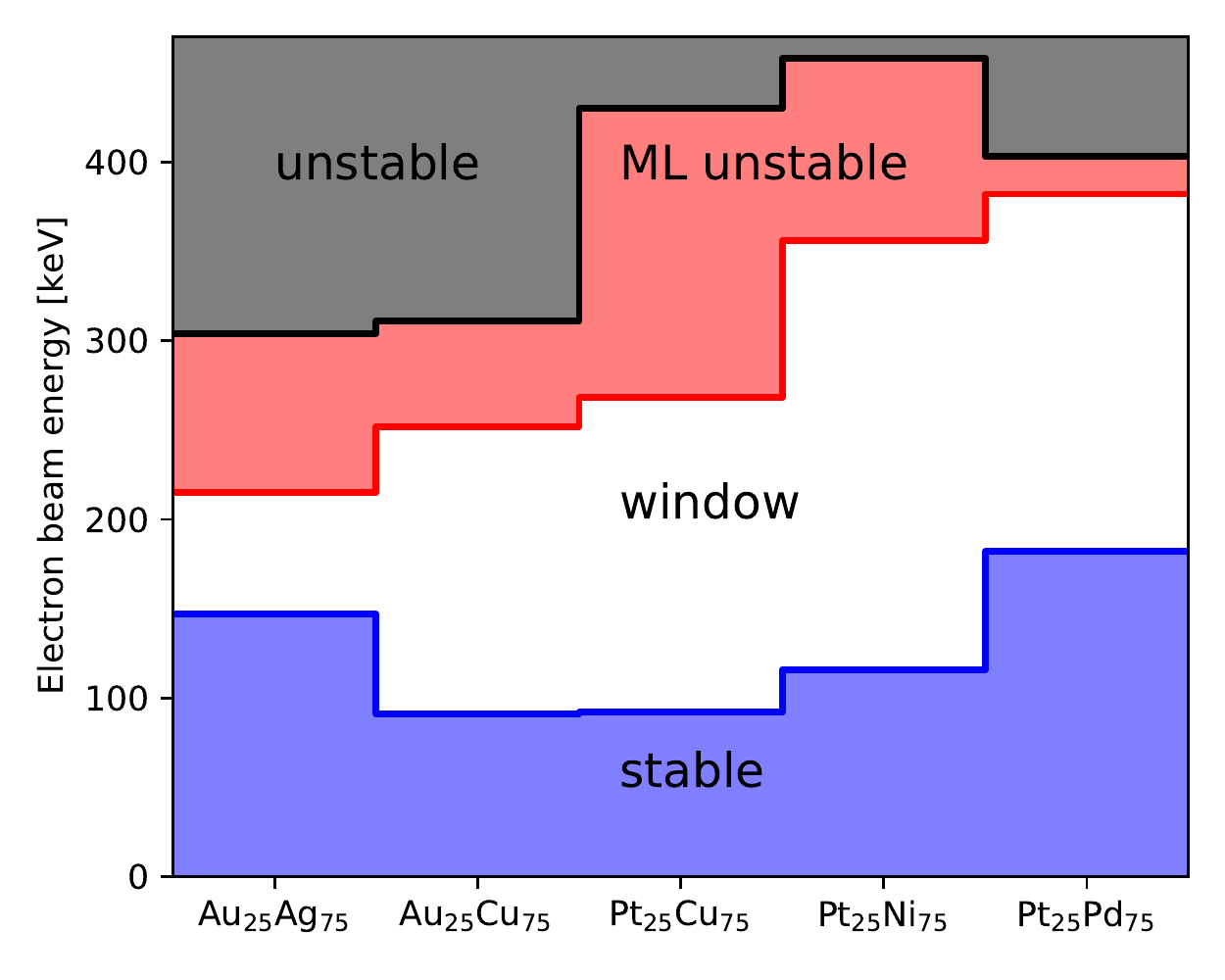}
\caption{Dealloying energy windows for five best alloy candidates. Shown are electron beam energy ranges [from Eq.~\eqref{eq:Emax}] that correspond to no dealloying (blue), only alloying secondary metal while keeping primary metal monolayer intact (white), dealloying secondary metal and primary metal monolayer (red), and dealloying all atoms (black).}
\label{fig:Emax}
\end{figure}
\end{center}

To identify promising alloy candidates for monolayer formation, let us now summarize our analysis. First, we exclude alloys with endothermic formation energies (negative numbers in Fig.~\ref{fig:histo}c). Second, we consider only alloys where the monolayer formation is kinetically feasible (positive numbers in Fig.~\ref{fig:DE}). Third, we exclude alloys where the alloyed monolayer at room temperature is unstable. These three considerations provide five promising alloy candidates (Fig.~\ref{fig:Emax}). With regard to the dealloying experiment, note that the beam energy required to remove Ag atoms from Au$_{25}$Ag$_{75}$ alloy is $E_\text{dealloy}=150$~keV while the Au monolayer can withstand $E_\text{stable}=220$~keV. The higher energy for Au atom removal is in qualitative agreement with the experiment.\cite{wang19} We also calculated the beam energies required to dealloy the secondary metal from the candidate binary alloys (Fig.~\ref{fig:Emax}). Dealloying Au atoms from Au$_{25}$Ag$_{75}$ alloy requires electron beam energy of 300 keV, which is twice the energy required to remove Ag atoms.

In conclusion, to investigate monolayer formation in alloy templates by electron irradiation, we performed a computational study on binary metal alloys composed of late d-series metals Ni, Cu, Pd, Ag, Pt, and Au. Our calculations show that removal of Ag atoms from Au$_{25}$Ag$_{75}$ alloy requires half of the electron beam energy required to remove Au atoms. Further, Au monolayer is stable against electron irradiation that is able to remove Ag atoms, in qualitative agreement with experiment. Our simulations suggest that Au monolayers could also be stable in Au--Cu alloy and Pt monolayers in Pt--Cu, Pt--Ni, and Pt--Pd alloys. Effects related to monolayer sizes deserve additional investigations, but we hope already these predictions will trigger new experiments to expand the family of free-standing 2D metals.

This work is supported by the Academy of Finland (Project No. 297115). The data that support the findings of this study are available from the corresponding author upon reasonable request.

%\section{}
%\label{}
%\subsection{}
%\subsubsection{}

% If in two-column mode, this environment will change to single-column format so that long equations can be displayed. 
%\begin{widetext}
%$$\mbox{put long equation here}$$
%\end{widetext}

% Figures should be put into the text as floats. 
% Use the graphics or graphicx packages (distributed with LaTeX2e).
% Put the label that you will use with \ref{} command in the braces of the \label{} command.
%
% \begin{figure}
% \end{figure}

% \begin{table}
% \end{tabular}
% \end{table}

%\begin{acknowledgments}
% Put your acknowledgments here.
%\end{acknowledgments}

\bibliography{your-bib-file}

\end{document}